\documentclass[12pt]{iopart}

\usepackage{iopams}  
\usepackage{graphicx}
\usepackage{bm}
\usepackage[dvips]{color}

\newcommand{\tc}{T_{2c}}
\newcommand{\tcp}{T_{2c^\prime}}

\begin{document}

\title{Discontinuous Transition of a Multistage Independent Cascade Model on Networks}

\author{Takehisa Hasegawa$^1$ and Koji Nemoto$^{2}$}
\address{$^1$ Department of Mathematical Informatics,
Graduate School of Information Science, Tohoku University, 6-3-09, Aramaki-Aza-Aoba, Sendai, Miyagi, 980-8579, JAPAN.
}
\address{$^2$ 
Department of Physics, Hokkaido University, Kita 10 Nishi 8, Kita-ku, Sapporo, Hokkaido, 060-0810, JAPAN
}
\ead{hasegawa@m.tohoku.ac.jp} 
\ead{nemoto@statphys.sci.hokudai.ac.jp}

\begin{abstract}
We propose a multistage version of the independent cascade model, 
which we call a multistage independent cascade (MIC) model, on networks.
This model is parameterized by two probabilities: 
the probability $T_1$ that a node adopting a fad increases the awareness of a neighboring susceptible node, 
and the probability $T_2$ that an adopter directly causes a susceptible node to adopt the fad.
We formulate a tree approximation for the MIC model 
on an uncorrelated network with an arbitrary degree distribution $p_k$.
Applied on a random regular network with degree $k=6$, this model exhibits a rich phase diagram, 
including continuous and discontinuous transition lines for fad percolation, 
and a continuous transition line for the percolation of susceptible nodes.
In particular, the percolation transition of fads is discontinuous (continuous) 
when $T_1$ is larger (smaller) than a certain value. 
A similar discontinuous transition is also observed in random graphs and scale-free networks. 
Furthermore, assigning a finite fraction of initial adopters dramatically changes the phase boundaries.
\end{abstract}


\maketitle


\section{Introduction}

Word-of-mouth phenomena often catapult books, movies, and music out of obscurity and into popularity.
Similar propagative effects are observed in voting behavior, the spread of rumors, and diffusion of innovations. 
Several models for describing such information cascades have been proposed in social science \cite{granovetter1978threshold,schelling1973hockey,valente1995network,mahajan1990new,bikhchandani1992theory}. 
In the context of network science, Watts proposed a network model (the linear threshold model \cite{granovetter1978threshold} on a network), 
in which the decision made by each node (person) to adopt an innovation is determined 
by the relative number of its adjacent nodes (friends) who have already adopted it \cite{watts2002simple,watts2007influentials}. 
As the decision threshold is lowered, the model transits from the local cascade phase, where cascades are constrained, 
to the global cascade phase, where some small initial shocks are propagated through the whole system. 
This transition is discontinuous; 
the final fraction of adopters rises discontinuously, and its distribution over many trials is bimodal even at the transition threshold 
\footnote{
Watts claimed that the linear threshold model on networks 
show both continuous (in a low connectivity region) and discontinuous (in a high connectivity region) transitions \cite{watts2002simple}. 
The continuous transition is essentially due to the percolation of underlying network, and
the model is expected to show only discontinuous transition if the network has sufficient connectivity.
}. 
Following the seminal paper by Watts, several authors have investigated the effects of network topology on the transition threshold~\cite{centola2007cascade,Gleeson2007seed,gleeson2008cascades,payne2009information,ikeda2010cascade,hackett2011cascades}.

The linear threshold model assumes {\it permanently active} nodes; that is, a node that has adopted an innovation will retain that innovation.
On the other hand, the susceptible-infected-removed (SIR) model \cite{anderson1992infectious} 
and the independent cascade (IC) model \cite{goldenberg2001using,kempe2003maximizing} 
(including the SIR model with transmissibility \cite{Newman-Spread-2002PRE}) have been investigated as models of {\it transient fads}. 
In the IC model, a fad adopter abandons the fad after a single attempt, with a certain probability, 
to transmit the fad to each susceptible neighbor. 
The SIR and IC models on complex networks have been extensively studied both analytically and numerically  
 \cite{Pastor-Satorras2001epidemicPRL,Pastor-Satorras2001epidemicPRE,Newman-Spread-2002PRE}. 
On networks, the transition of these models is not discontinuous but continuous
\footnote{
The SIR model is mapped onto the IC model with a transmission probability (the SIR model with transmissibility). 
Although these two behave differently from each other on local scales, 
there is a correspondence between these two for critical points and critical exponents 
\cite{tome2010critical,kenah2007second,miller2007epidemic,kenah2011epidemic}.
}.

Several complex contagion models have been recently reported \cite{dodds2004universal,antal2012outbreak,kerchove2009role,melnik2013multi,krapivsky2011reinforcement,starnini2012ordering}.
Even transient fads may undergo discontinuous transition if more than one state is available to each node before adoption. 
Incorporating {\it social reinforcement} into the SIR model, 
Krapivsky et al. developed an agent-based model of transient fads, which we call the fad model \cite{krapivsky2011reinforcement}. 
Here, social reinforcement means that each node adopts a fad only after multiple prompts from adjacent adopters.
In this model, a node takes one of the $M+1$ awareness levels $(0, 1, \cdots$, $M)$: 
Nodes with the awareness level less than $M$ remain susceptible, and nodes with the highest awareness level $M$ adopt the fad. 
Each adopter increases the awareness level of its adjacent neighbors by one at a unit rate. 
Adopters also abandon the fad at a certain rate.
In the absence of reinforcement ($M=1$), the model reduces to an ordinary SIR model.
Krapivsky et al.\ studied this model without a spatial structure by the macroscopic rate equations \cite{krapivsky2011reinforcement}. 
They showed that the transition is continuous for $M = 1$ and discontinuous for $M \ge 2$. 
However, the behavior of this agent-based model on a network remains unknown.

In this study, we analytically investigate the effect of social reinforcement on transient fads in networked systems. 
To this end, we propose a simple multistage independent cascade (MIC) model, a multistage version of the IC model.
The present MIC model is a discrete time fad model of $M=2$. 
Provided that the underlying network is locally treelike, the tree approximation is applicable to this model.
Here, we formulate the tree approximation for uncorrelated networks of arbitrary degree distribution $p_k$.
As an application, we study this model on a random regular network of degree $k = 6$, random graphs, and scale-free networks, 
to show that 
the present model actually exhibits a discontinuous jump of the number of abandoners. 
We also investigate the percolation transitions in networks of susceptible nodes and abandoners.
Interestingly, when the fraction of initial adopters is finite, 
the number of abandoners can discontinuously jump after the abandoner's network has percolated.
Overall, the phase diagram of the MIC model crucially depends on the fraction of initial fad adopters.

\begin{figure}
 \begin{center}
  \includegraphics[width=40mm]{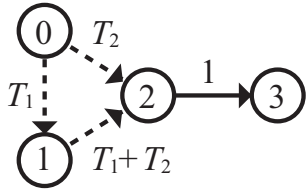}
   \end{center}
 \caption{
Transition rules of the present MIC model. 
The values on the lines indicate the transmission probability.
The events indicated by the dashed lines ($0 \to 1, 0 \to 2$, and $1 \to 2$) occur at nodes adjacent to an adopter
The event indicated by the solid line ($2 \to 3$) occurs irrespective of the neighbor's state.
}
 \label{rule}
\end{figure}

\section{Model}

Consider a network of $N$ nodes. 
In our MIC model, each node takes one of the four states: (0) susceptible and unaware, 
(1) informed (susceptible and aware), (2) fad adopter, and (3) fad abandoner.
The model dynamics is given as follows (figure~\ref{rule}):
\begin{itemize}
\item[(i)] 
Randomly select a fraction $\rho$ of nodes as the initial fad adopters ({\em seeds}). Set the state of other nodes to susceptible.
\item[(ii)] 
Compile a (randomly ordered) list $L$ of adopters. 
For each adopter (2) in $L$, compile a list $S$ of the neighboring susceptible nodes (0 or 1). For each node in $S$, 
execute the following process: (ii-a) If the node is susceptible and unaware (0), 
change it to either informed ($0 \to 1$) with probability $T_1$, 
fad adopter ($0 \to 2$) with probability $T_2$, or leave unchanged otherwise. 
(ii-b) If the node is informed (1), change it to fad adopter ($1 \to 2$) 
with probability $T\equiv T_1+T_2 (\le 1)$, or leave unchanged.
At this stage, the new adopters are not yet included in $L$.
\item[(iii)] 
Change all adopters in $L$ to abandoners ($2 \to 3$). 
\item[(iv)] 
Repeat (ii) and (iii) until no adopter exists in the network.
\end{itemize}
Note that when $T_1=0$, the model reduces to the IC model (the SIR model with transmissibility $T_2$).

In the following sections, we denote the final state fraction of nodes in states 0, 1, 2, and 3 by $S_0$, $S_1$, $S_2(=0)$, and $S_3$, respectively. 
We also designate the networks of susceptible nodes and abandoners as the $S_0$-network and $S_3$-network, respectively.
The fractions of the largest components of the $S_0$- and $S_3$-networks are, respectively, denoted by $S_0^{\rm max}$ and $S_3^{\rm max}$.

\section{Phase Diagram in the $\rho \to 0$ limit}

In this section, we consider the dynamics starting from an infinitesimal fraction of seeds ($\rho = 0 +$) 
on uncorrelated networks with a given degree distribution $p_k$ 
(for example, a random regular network, Erd{\H{o}}s--R{\'e}nyi random graph~\cite{erdos1960evolution}, 
or a scale-free network (SFN) generated by using the configuration model~\cite{molloy1995critical}). 
Newman analyzed the SIR model with transmissibility $T$ (the present model with $T_1=0$ and $T_2=T$) 
using generating functions \cite{Newman-Spread-2002PRE,Newman-Threshold-2005PRL}.
He reported two phase transitions as $T$ is increased:
a giant component of abandoners appears at the percolation threshold of the $S_3$-network $T_c = \langle k\rangle/\langle k^2-k\rangle$ 
(where $\langle \cdot \rangle$ denotes the average of a quantity weighted by $p_k$), 
and a giant component of susceptible nodes disappears at the percolation threshold of the $S_0$-network $T_{c'}$. 
Correspondingly, there exist three distinct phases: 
$S_0^{\max} >0$ and $S_3^{\max}=0$ for $T<T_c$, 
$S_0^{\max}>0$ and $S_3^{\max}>0$ for $T_c < T < T_{c'}$, and $S_0^{\max}=0$ and $S_3^{\max}>0$ for $T>T_{c'}$.
In the following, we generalize his method in order to apply it to the present model. 

\subsection{Percolation of the $S_3$-network}

We investigate the phase diagram of percolation (of the $S_0$- and $S_3$-networks), 
and therefore focus on the final state of the dynamics, where no adopters exist.
Consider a randomly chosen (RC) node and its RC neighbor.
Here we assume that the focal RC node was still susceptible or informed when the RC neighbor became an adopter (if such a neighbor exists).
We denote the probability that 
the prompt of this RC neighbor did not change the state of the focal node by $u$,
and the probability that the RC neighbor is susceptible or informed by $v$ in the final state. 
Then $u$ and $v$ are related as
\begin{equation}
u = v + (1-T)(1-v) = 1-T+Tv. \label{self-SIR}
\end{equation}
The first term $v$ on the right-hand side of the first equality signifies that 
the RC neighbor did not adopt the fad and that no transmission occurred from this RC neighbor to the focal node. 
The second term $(1-T)(1-v)$ is the probability that 
the RC neighbor abandons the fad, but has never altered the state of the focal node while adopting the fad.

We can evaluate the fractions of each state, $S_0$, $S_1$, and $S_3$, by $u$ and $v$.
To this end, it is convenient to introduce the generating function of the degree distribution $p_k$ and the excess degree distribution $q_k$: 
\begin{eqnarray}
F_0(x)&=&\sum_k p_k x^k, \quad F_1(x)= \sum_k q_k x^k.
\end{eqnarray}
Here, $q_k$ is the probability that an endpoint of an edge (an RC neighbor of an RC node) has excess degree $k$, i.e., 
the number of neighbors minus one is $k$. In uncorrelated networks, 
$q_k=(k+1) p_{k+1}/\langle k \rangle$, so that $F_1(x)=F_0^\prime(x)/F_0^\prime(1)$.
Because the probability that an RC node with degree $k$ is susceptible is $u^k$, 
the fraction of susceptible nodes $S_0$, which is equal to the probability that an RC node is susceptible, is
\begin{equation}
S_0=F_0(u)=\sum_k p_k u^k.
\end{equation}
A node with degree $k$ remained informed if it received information from just one of the adopters among its $k$ neighbors 
and it did not receive any information from other neighbors.
The probability that an RC neighbor was an adopter is $1-v$ 
and an informing attempt succeeds with probability $T_1$. 
Therefore the probability that an RC node is informed is $\sum_k p_k {k \choose 1} u^{k-1} (1-v) T_1 =T_1(1-v)F_0'(u)$.
Thus we obtain
\begin{equation}
S_1=T_1(1-v)F_0'(u). \label{S1eq}
\end{equation}
Since $S_2=0$ in the final state, we have 
\begin{equation}
S_3=1-S_0-S_1.
\end{equation}

To obtain a self-consistent equation for $u$, we divide $v$ into two parts, $v=v_0+v_1$, 
where $v_0$ and $v_1$ are the probabilities that an RC neighbor of an RC node is susceptible and informed, respectively.
Because $u^k$ is the probability that an RC neighbor with excess degree $k$ is susceptible 
and $q_k$ is the excess degree distribution, we have 
\begin{equation}
v_0 =F_1(u)=\sum_k q_k u^k.
\end{equation}
Because the probability that an RC neighbor with excess degree $k$ is informed is 
${k \choose 1} u^{k-1} (1-v) T_1$, we have
\begin{equation}
v_1 =T_1(1-v)F_1'(u).
\end{equation}
Adding $v_0$ and $v_1$, we obtain 
\begin{equation}
v = F_1(u)+T_1(1-v)F_1'(u). \label{eqv}
\end{equation}
Equations~(\ref{self-SIR}) and~(\ref{eqv}) provide a self-consistent equation for $u$, 
\begin{equation}
u=1-T+TF_1(u)+T_1(1-u)F_1'(u).
\label{sceFAD}
\end{equation}

We now analyze in detail the solution of the self-consistent equation.
For any value of $T_1$ and $T_2$, there exists a trivial solution $u=1$, implying that $S_0=1$, $S_1=0$, and thus $S_3=0$.
To find nontrivial solutions, we solve equation~(\ref{sceFAD}) with respect to $T_2$:
\begin{equation}
T_2=\varphi_0(u;T_1)\equiv(1-u)\frac{1+T_1F_1'(u)}{1-F_1(u)}-T_1.
\label{sceFAD1}
\end{equation}
This gives $T_2$ such that a given value of $u<1$ is a solution of equation~(\ref{sceFAD}) with a fixed $T_1$.
For small $\bar u=1-u \ll 1$, equation~(\ref{sceFAD1}) can be expanded to give
\begin{equation}
T_2=\frac{1}{c_1}+\frac{c_2(1-c_1T_1)}{2c_1^2}\bar u 
+\frac{(3c_2^2-2c_1c_3)+c_1(4c_1c_3-3c_2^2)T_1}{12 c_1^3}\bar u^2 + \cdots,
\label{sceFAD2}
\end{equation}
where $c_n=F_1^{(n)}(1)$.
Taking the limit $u \to 1-$ ($\bar u \to 0+$), we have 
\begin{equation}
T_{2}=\tc \equiv c_1^{-1}=\frac{1}{F_1'(1)}=\frac{\langle k\rangle}{\langle k^2-k\rangle},
\end{equation}
above which the trivial solution $u=1$ becomes unstable and $S_3$ is nonzero, 
implying that $S_3^{\rm max}$ changes from $S_3^{\rm max}=0$ to $S_3^{\rm max}>0$.

The type of transition at $T_2=\tc$ depends on the value of $T_1$.
For fixed $T_1<\tc$, a nontrivial solution exists only for $T_2>\tc$ and $S_3\to 0$ when $T_2\to \tc+$. 
Therefore $T_2=\tc$ is the continuous transition point of $S_3$.
For $T_1>\tc$, on the other hand, $u<1$ is possible even when $T_2<\tc$.
To see this, we consider small $\Delta_1=T_1-\tc$ and $\Delta_2=\tc-T_2$ and let us rewrite equation~(\ref{sceFAD2}) for 
\begin{equation}
\Delta_2=2a\Delta_1\bar u -b\bar u^2 + \cdots, 
\end{equation}
where $a$ and $b$ are positive constants.
When $\Delta_1>0$, $\Delta_2$ can be positive for small $\bar u>0$, yielding two branches of the nontrivial solution 
for $T_d^L<T_2<T_d^U$ with
\begin{equation}
T_d^L\simeq \tc-\frac{a^2}{b}\Delta_1^2, \quad T_d^U=\tc
\end{equation}
and the lower branch vanishes at $T_d^U$ while the upper branch, which gives a positive $S_3$, survives above $T_d^U$.
By contrast, when $T_2<T_d^L$, the trivial solution alone exists and the only possibility is $S_3=0$.
This implies that $S_3$ {\em discontinuously} jumps from zero to the upper branch within the range $T_d^L\le T_2\le T_d^U$.
In general, $T_d^L$ depends on $T_1$ and is given by $\displaystyle T_d^L(T_1)=\min_{0<u<1}\varphi_0(u;T_1)$.

\subsection{Percolation of $S_0$-network}

When $T_2$ increases with $T_1$ fixed, 
the $S_0$-network gradually shrinks and finally disintegrates into numerous finite components at $T_2=\tcp$.
Here, we examine this percolation problem.

Consider an RC node and its RC neighbor.
The probabilities that the RC neighbor is susceptible, informed, and an abandoner that failed 
to affect the focal node while an adopter are $v_0$, $v_1$, and $(1-T)(1-v)$, respectively.
The probability that a node in the $S_0$-network has $m$ susceptible neighbors, 
i.e., the degree distribution $\pi_m$ of the $S_0$-network, is represented as
$\pi_m \propto \sum_{k=m}^\infty p_k {k \choose m} {v_0}^m [v_1+(1-T)(1-v)]^{k-m}$. 
The corresponding generating function is
\begin{equation}
G_0(x) 
=\sum_m\pi_m x^m
=\frac{F_0\left[v_0x+v_1+(1-T)(1-v)\right]}{F_0(u)}
=\frac{F_0\left[v_0(x-1)+u\right]}{F_0(u)}.
\end{equation}
Here the denominator is the prior probability of being susceptible.
Similarly, the generating function for the excess degree distribution of the $S_0$-network is derived as
\begin{equation}
G_1(x)=\frac{G_0'(x)}{G_0'(1)}=\frac{F_1\left[v_0(x-1)+u\right]}{F_1(u)}.
\end{equation}
Because the mean excess degree of the $S_0$-network is given by $G_1'(1)$, 
$u$ at the percolation threshold $\tcp$ (for a given $T_1$) satisfies
\begin{equation}
1=G_1'(1)=v_0\frac{F_1'(u)}{F_1(u)}=F_1'(u) \label{s0threshold}.
\end{equation}
By following \cite{Newman-Threshold-2005PRL}, we have the largest component of the $S_0$-network. 
The fraction of the largest component in the $S_0$-network, $C$, is 
\begin{equation}
C=1-G_0(v_S),
\end{equation}
where $v_S$ is the probability that an $S_0$-component connected by an RC edge is finite,
\begin{equation}
v_S=G_1(v_S).
\end{equation}
Then the fraction of the largest component of the $S_0$-network over the {\it whole network}, i.e., $S_0^{\rm max}$, is given as 
\begin{equation}
S_0^{\rm max} =C S_0 = S_0 (1-G_0(v_S)). \label{s0max}
\end{equation}

\begin{figure}
 \begin{center}
  \includegraphics[width=75mm]{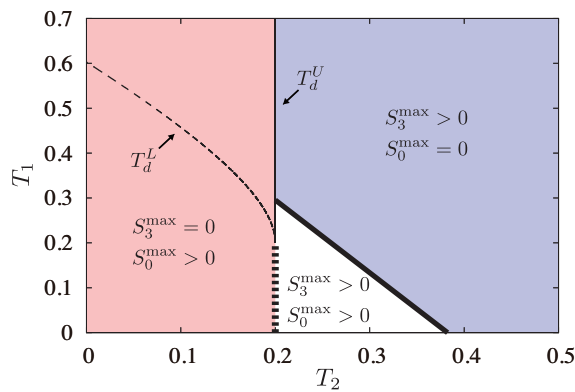}
   \end{center}
 \caption{
Phase diagram for the case $\rho = 0+$. 
The thick-solid and thick-dashed lines denote the continuous transition lines for the percolation of the $S_0$-network $\tcp$ 
and the $S_3$-network $\tc$, respectively.
The thin-solid and thin-dashed lines denote the upper ($T_d^U$) and lower ($T_d^L$) bounds, respectively.
Between these bounds, equation~(\ref{sceFAD}) has two nontrivial solutions.
}
 \label{fig-phasediagramrho0}
\end{figure}

\begin{figure}
 \begin{center}
  \includegraphics[width=75mm]{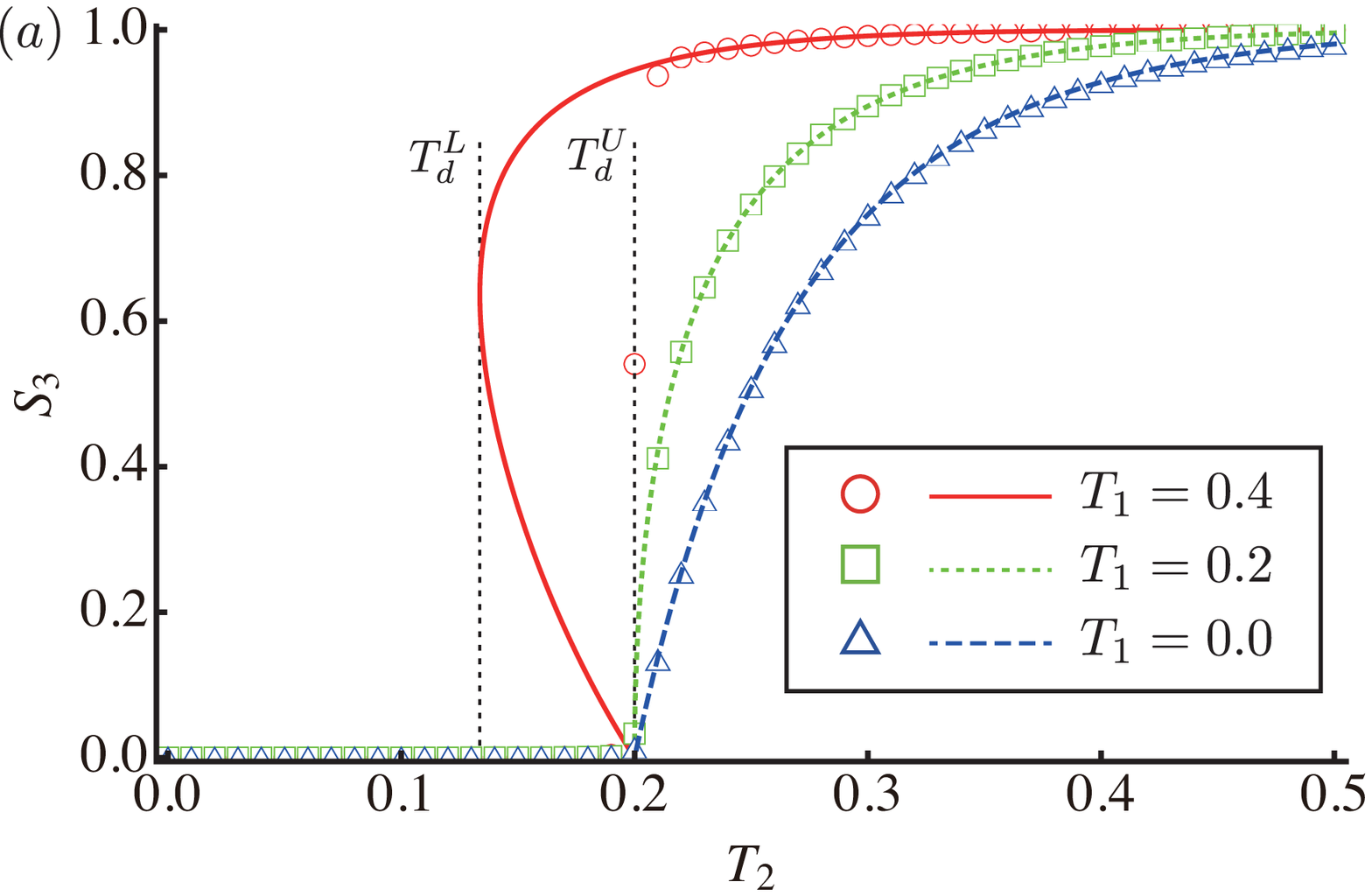}
  \includegraphics[width=75mm]{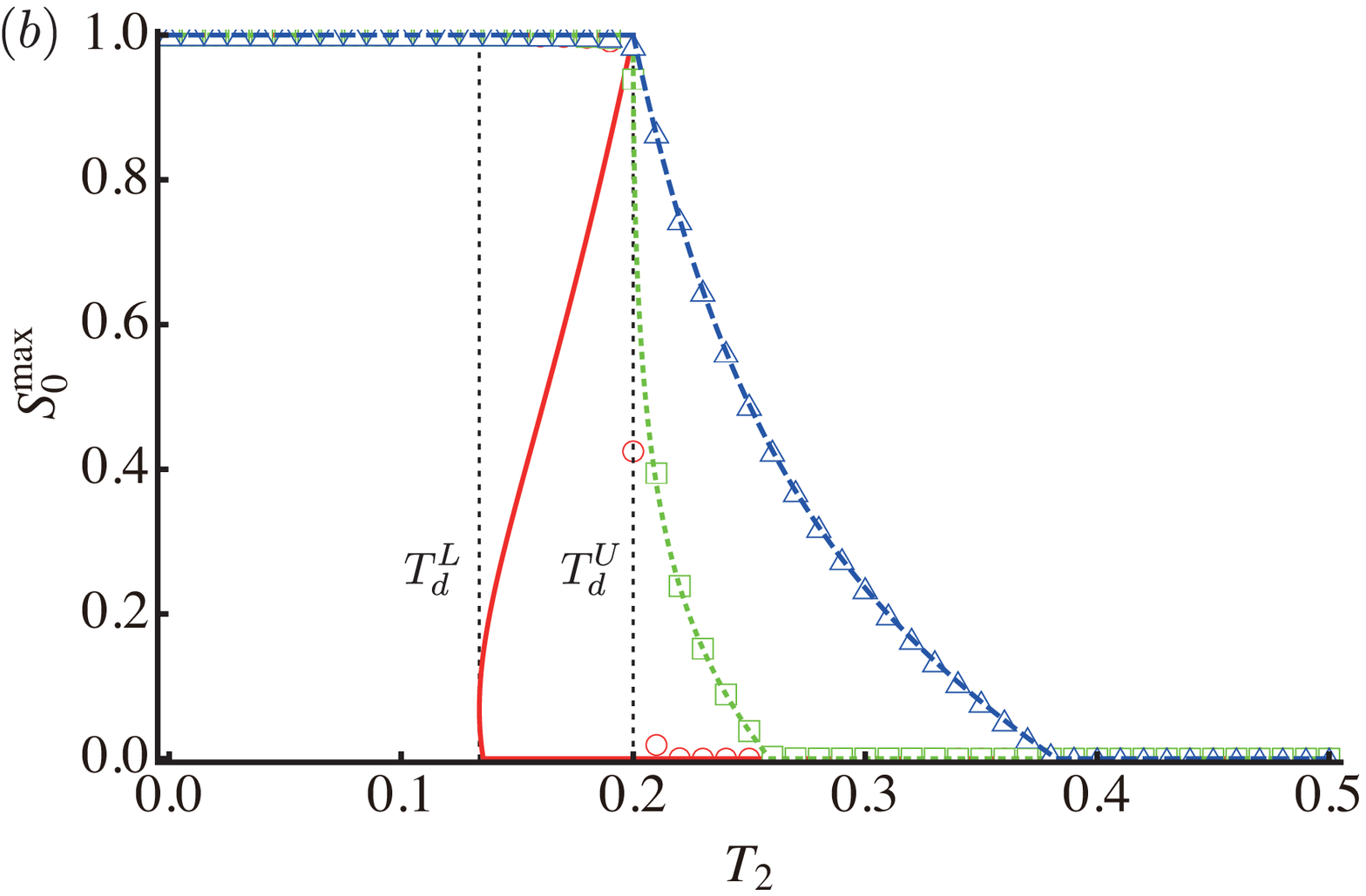}
   \end{center}
 \caption{
Analytical results (lines) for (a) $S_3$ and (b) $S_0^{\rm max}$ as a function of $T_2$ with $T_1=0$, $0.2$, and $0.4$.
$T_d^L$ and $T_d^U$ indicate the lower and upper bounds, respectively. 
Between these bounds, equation~(\ref{sceFAD}) has two nontrivial solutions.
Symbols are obtained by Monte-Carlo simulations with very small $\rho (=0.0001)$.
The number of nodes used is $N=256000$, and 100 trials for each of 100 graph realizations are used for averaging.
}
 \label{fig-analyticalS3rho0}
\end{figure}

\begin{figure}
 \begin{center}
  \includegraphics[width=75mm]{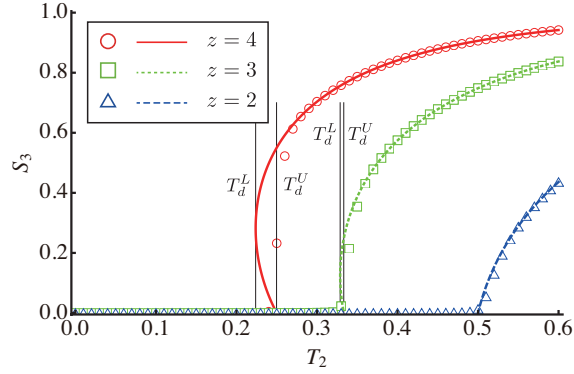}
   \end{center}
 \caption{
Analytical result (lines) for $S_3$ of the random graph, as a function of $T_2$, for $T_1=0.4$ and several values of average degree $z$.
Symbols are obtained by Monte-Carlo simulations with very small $\rho (=0.0001)$.
The number of nodes used is $N=256000$, and 100 trials for each of 100 graph realizations are used for averaging.
}
 \label{fig-analyticalS3-ER}
\end{figure}

\begin{figure}
 \begin{center}
  \includegraphics[width=50mm]{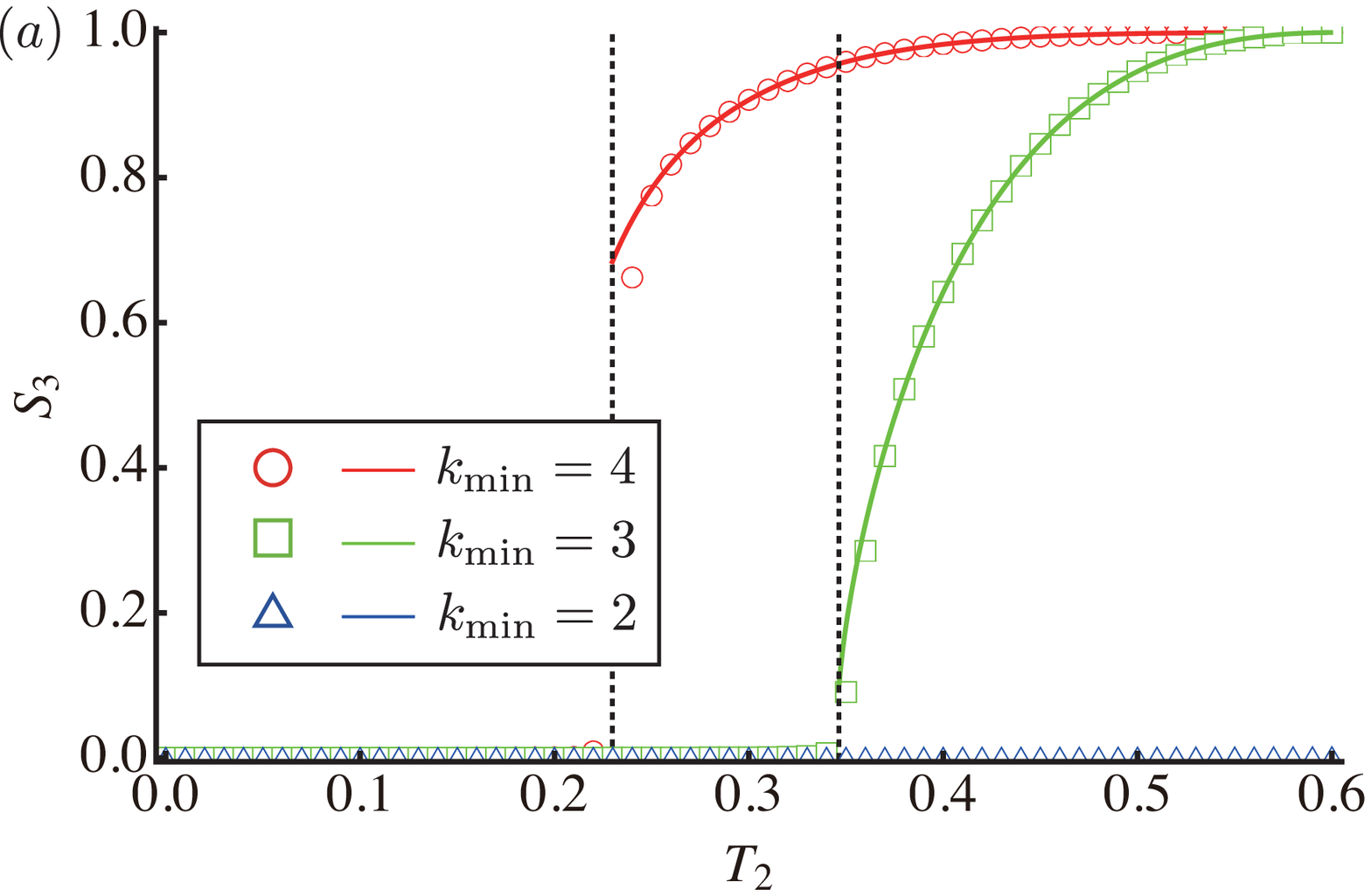}
  \includegraphics[width=50mm]{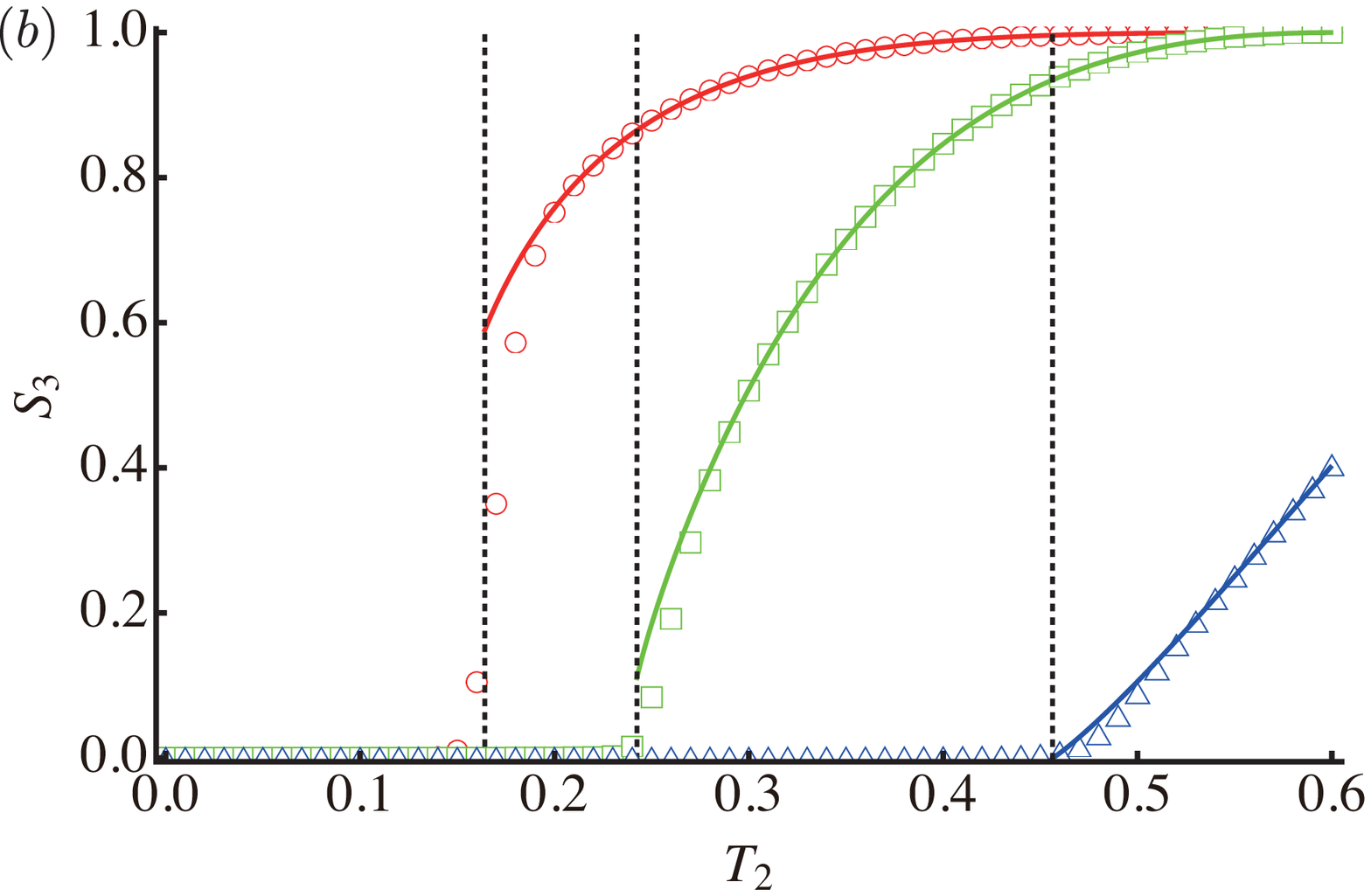}
  \includegraphics[width=50mm]{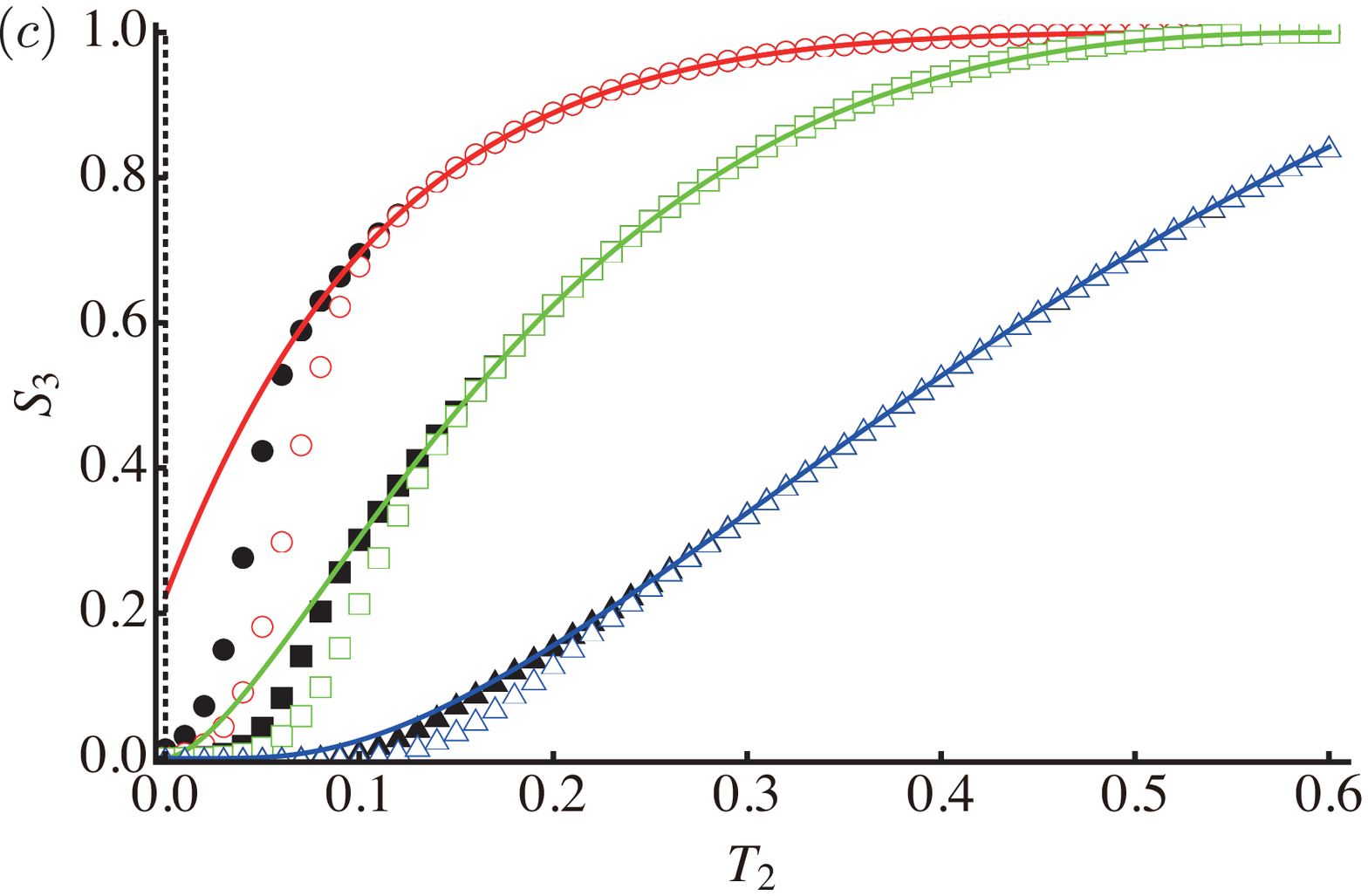}
   \end{center}
 \caption{
Analytical result (lines) for $S_3$ of the uncorrelated SFN with $T_1=0.4$ and several values of $k_{\rm min}$, as a function of $T_2$.
In each panel, the degree exponent $\gamma$ is set to (a) $\gamma=$5, (b) $\gamma=$4, and (c) $\gamma=$3, respectively.
The vertical lines represent $\tc$.
Open symbols are the numerical results of $S_3$ obtained by Monte-Carlo simulations with very small $\rho (=0.0001)$.
The number of nodes used is $N=256000$, and 100 trials for each of 100 graph realizations are used for averaging.
The full symbols in (c) denote $S_3$ obtained by the Monte-Carlo simulations with a larger system size ($N=1024000$).
}
 \label{fig-SFN}
\end{figure}


\subsection{Phase diagram for random regular network}

We now consider the phase diagram of the present model on a degree-6 random regular network.
Figure~\ref{fig-phasediagramrho0} shows the phase diagram in $T_1$--$T_2$ space obtained by applying the tree approximation.
The transition line of the $S_3$-percolation is independent of $T_1$ and locates at $T_2=\tc=1/(k-1)=0.2$.
However, the order of the transition is continuous when $T_1<\tc$, and otherwise discontinuous.
In figure~\ref{fig-analyticalS3rho0} (a), we plot $S_3$ as a function of $T_2$ for several values of $T_1$. 
When $T_1$ is large, $S_3$ is multivalued in $T_d^L<T_2<T_d^U=\tc$. 
Since the dynamics begins from an infinitesimal fraction of seeds, $S_3$ remains at the lower branch for $T_2<T_d^U$ 
and jumps to the upper branch at $T_d^U$, 
implying that a discontinuous occurrence of the giant component of the $S_3$-network.
Such a jump disappears when $T_1<\tc$. 
In figure~\ref{fig-analyticalS3rho0}(a), we also plot the numerical results obtained by the Monte-Carlo simulations.
Our Monte-Carlo results with very small $\rho(=0.0001)$ correspond well with the analytical calculations. 
The $N$- and $\rho$-dependence of the Monte-Carlo results will be checked later in figures~\ref{fig-comparisonR} and~\ref{fig-analyticR}, respectively.

For small $T_1$ (the blue-triangles and green-squares in figure~\ref{fig-analyticalS3rho0}(b)), 
giant components of the $S_0$- and $S_3$-networks coexist in $\tc<T_2<\tcp$. 
This region shrinks with increasing $T_1$, and $\tcp(T_1)$ reaches $\tc$ at a certain value of $T_1$ ($T_1 \simeq 0.295$).
Above this value (the red-circles in figure~\ref{fig-analyticalS3rho0}(b)), the $S_3$-network percolates and the giant component of the $S_0$-network simultaneously vanishes at $T_2=\tc$.
The tricritical point $(\tc, \tc)$ of the random regular network with degree $k$ is given by $(1/(k-1), 1/(k-1))$. 
We note that $(\tc, \tc)=(0.5,0.5)$ when $k=3$, precluding $T_1>\tc$ at $T_2=\tc$ since $T=T_1+T_2\le 1$.
Consequently, the present model on a degree-3 random regular network exhibits only a continuous transition of the $S_3$-network; discontinuous transitions are disallowed.

We now remark on other typical locally treelike networks; namely, 
the random graph with $p_k = e^{-z}z^k/k!$ and the uncorrelated SFN 
with $p_k=k^{-\gamma}/\sum_{k=k_{\rm min}}k^{-\gamma}$, where $k_{\rm min}$ is the minimum degree. 

For random graphs with average degree $z$, the generating functions are given as $F_0(x)=F_1(x)=\exp[z(x-1)]$, 
and the tricritical point $(\tc, \tc)$ is given by $(1/z, 1/z)$. 
Thus, the MIC model undergoes a discontinuous transition when $z > 2$ rather than when $z > 3$.
Figure~\ref{fig-analyticalS3-ER} shows the analytical and numerical results for $S_3$ as a function of $T_2$ with $T_1=0.4$ and $z=$4, 3, and 2. 
Numerical results support the validity of our tree approximation, while it is hard to assess numerically the discontinuity of $S_3$ 
when $z=3$ because the interval of $T_d^L<T_2<T_d^U$ is very narrow.

Figure~\ref{fig-SFN} shows the analytical and numerical results for $S_3$ 
of the uncorrelated SFN with $T_1=0.4$ and several values of $\gamma$ and $k_{\rm min}$. 
For the case of $\gamma=5$, $S_3$ discontinuously jumps at $\tc$ 
when $k_{\rm min}=3,4$ (although the discontinuity at $k_{\rm min}=3$ is weak as well as the random graph with $z=3$), 
but $S_3$ is always zero when $k_{\rm min}=2$, where $\tc=1/F_1'(1)>0.6$.
For the case of $\gamma=4$, $S_3$ can be nonzero even when $k_{\rm min}=2$. 
But the transition at $\tc$ is continuous one because $\tc \simeq 0.456 > T_1(=0.4)$.
We have a discontinuous jump of $S_3$ when $k_{\rm min}=3,4$.
For the case of $\gamma=3$, $\tc=0$ because $F_1'(1)=\langle k^2 -k \rangle/\langle k \rangle$ diverges irrespectively of $k_{\rm min}$. 
Therefore, both continuous and discontinuous transition lines of the $S_3$-network disappear, 
and only the percolation transition of the $S_0$-network exists.
Monte-Carlo simulations for the case of $\gamma=3$ suffer a finite size effect in a small $T_2$-region, 
although the deviation from the analytical lines will vanish with increasing $N$ (see the full symbols in figure \ref{fig-SFN}(c)).

\section{Sensitivity to the fraction of initial adopters}

\subsection{Tree approximation}

From the formal analysis in the previous section, 
we can easily derive the self-consistent equations when the fraction of initial adopters $\rho > 0$.
Let $\bar\rho=1-\rho$ be the initial fraction of susceptible nodes.
In the present model, $v=v_0+v_1$ where
\begin{equation}
v_0=\bar\rho F_1(u),\quad v_1=\bar\rho T_1(1-v) F_1'(u),
\label{v0v1rho}
\end{equation}
and the relationship $u=1-T+Tv$ is unchanged. 
Thus, we obtain an expression in $u$ alone: 
\begin{equation}
u=1-T\left(1-\bar\rho F_1(u)\right) + \bar\rho T_1(1-u) F_1'(u). \label{u-eq-rhoneq0}
\end{equation}
By using the solution, we obtain
\begin{equation}
S_0=\bar\rho F_0(u),\quad
S_1=\bar\rho T_1(1-v)F_0'(u)=\bar\rho \frac{T_1(1-u) }{T}F_0'(u),
\label{S0S1rho}
\end{equation}
and $S_3=1-S_0-S_1$.

When $\rho>0$, there is no trivial solution of equation~(\ref{u-eq-rhoneq0}) for all regions of parameters and no continuous transition 
in the sense that $S_3$ is always nonzero because $S_3\ge \rho>0$. 
Nevertheless, the percolation of the $S_3$-network (from $S_3^{\rm max}=0$ to $S_3^{\rm max}>0$) 
can take place as long as $\rho$ is smaller than the percolation critical density (as shown in the next subsection). 
Moreover, a discontinuous jump of $S_3$ is still possible.
Equation (\ref{sceFAD1}) is modified to
\begin{equation}
T_2=\varphi_\rho(u; T_1)=(1-u)\frac{1+\bar\rho T_1 F_1'(u)}{1-\bar\rho F_1(u)}-T_1,
\end{equation}
and the lower and upper bounds of the discontinuous point ($T_d^L$ and $T_d^U$, respectively) 
can be given by the local minimum and local maximum of $\varphi_\rho(u; T_1)$ in $0<u<1$, respectively.
Depending on $T_1$, there exists some $\rho$, above which $\varphi_\rho(u; T_1)$ is monotonically decreasing and these bounds do not exist (figure~\ref{fig-criticalrho}).

On the $S_0$-network, the percolation threshold $\tcp(T_1)$ 
and the ratio of the largest component $S_0^{\max}$ are derived in the same manner; that is, 
(\ref{s0threshold}) and (\ref{s0max}) remain valid, along with (\ref{v0v1rho}) and (\ref{S0S1rho}).

\begin{figure}
 \begin{center}
  \includegraphics[width=75mm]{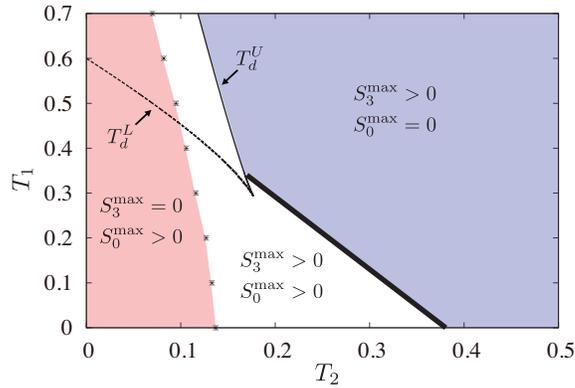}
   \end{center}
 \caption{
Phase diagram for the case $\rho =0.01$. 
The thin-solid and thin-dashed lines denote the upper ($T_d^U$) and lower ($T_d^L$) bounds, respectively.
Between these bounds, equation~(\ref{u-eq-rhoneq0}) has two nontrivial solutions.
The thick-dashed line denotes the transition line of percolation of the $S_0$-network $\tcp$.
Discrete dots denote the percolation transition points of the $S_3$-network obtained by Monte--Carlo simulations.
}
\label{fig-phasediagramhroneq0}
\end{figure}

\begin{figure}
 \begin{center}
  \includegraphics[width=75mm]{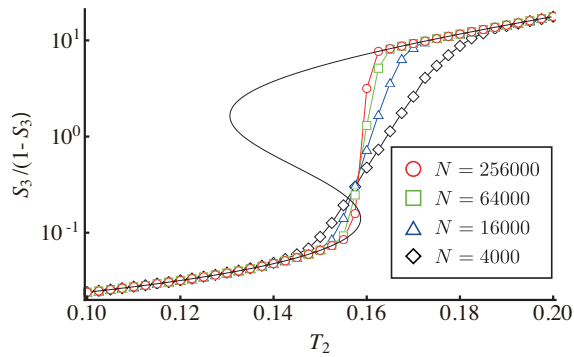}
   \end{center}
 \caption{
$S_3$ as a function of $T_2$ for $T_1=0.4$ and $\rho=0.01$. 
The analytic result (solid curve) and the result obtained by simulation are shown.
Here, $S_3$ is rescaled to $S_3/(1-S_3)$ for observing the behaviors near $S_3=0$ and 1.
The analytical evaluation gives $T_d^L\simeq 0.131$ and $T_d^U\simeq 0.159$. 
}
 \label{fig-comparisonR}
\end{figure}

\begin{figure}
 \begin{center}
  \includegraphics[width=75mm]{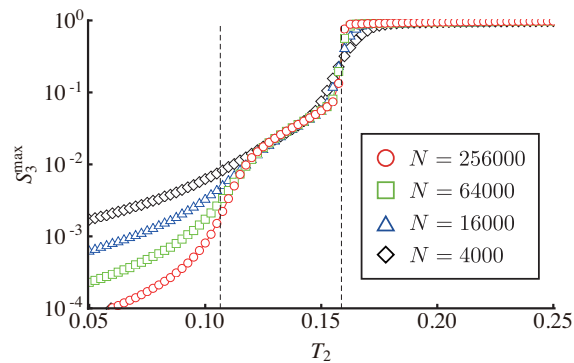}
 \end{center}
 \caption{
Monte--Carlo results of $S_3^{\rm max}$ as a function of $T_2$, with $T_1=0.4$ and $\rho=0.01$. 
The two vertical lines represent $\tc \simeq 0.106$ and $T_d^U \simeq 0.159$.
$\tc \simeq 0.106$ is given by the cross point of the effective fractal exponents of differently sized systems.
}
\label{fig-Rmax}
\end{figure}

\begin{figure}
 \begin{center}
  \includegraphics[width=75mm]{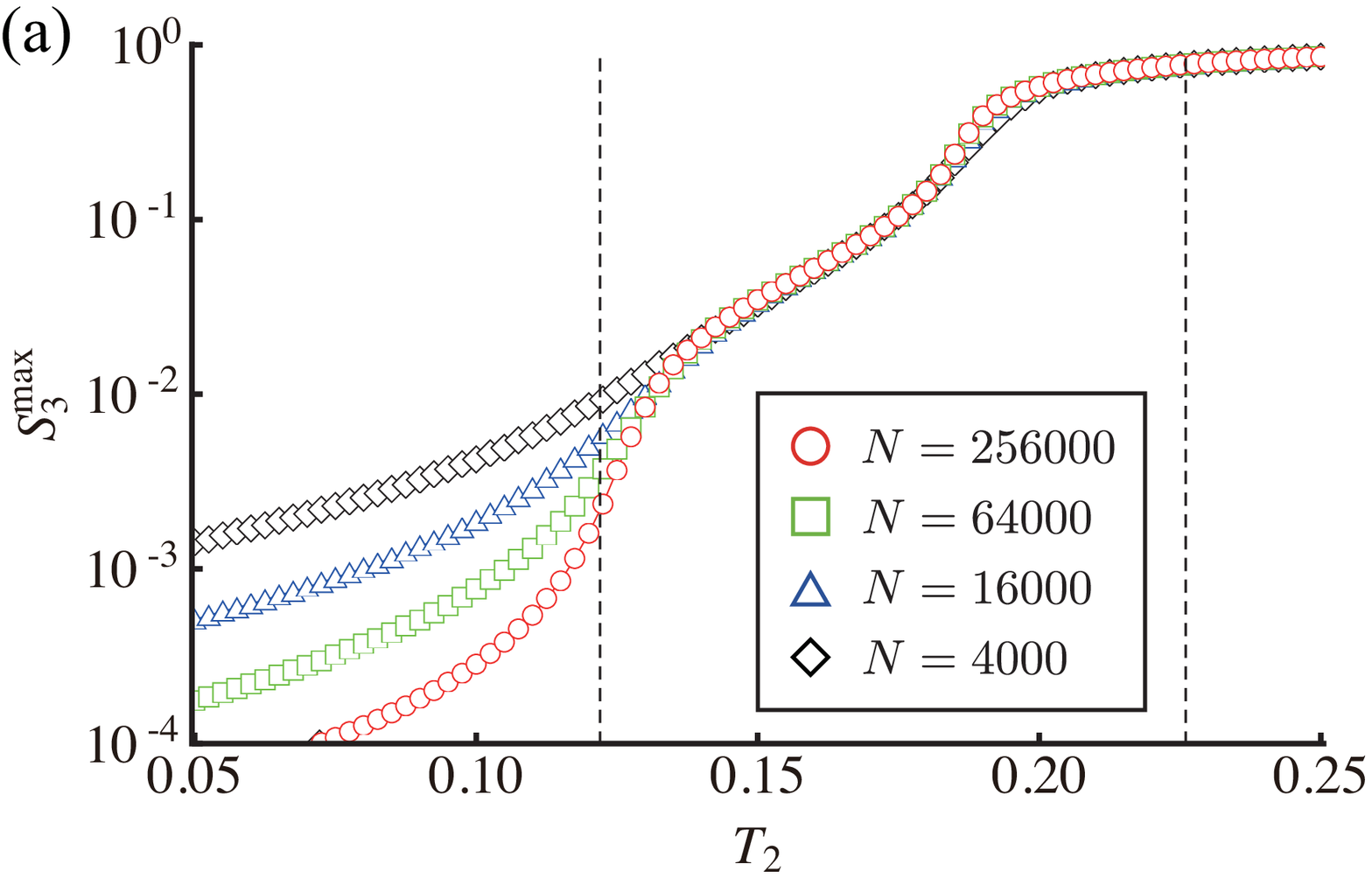}
  \includegraphics[width=75mm]{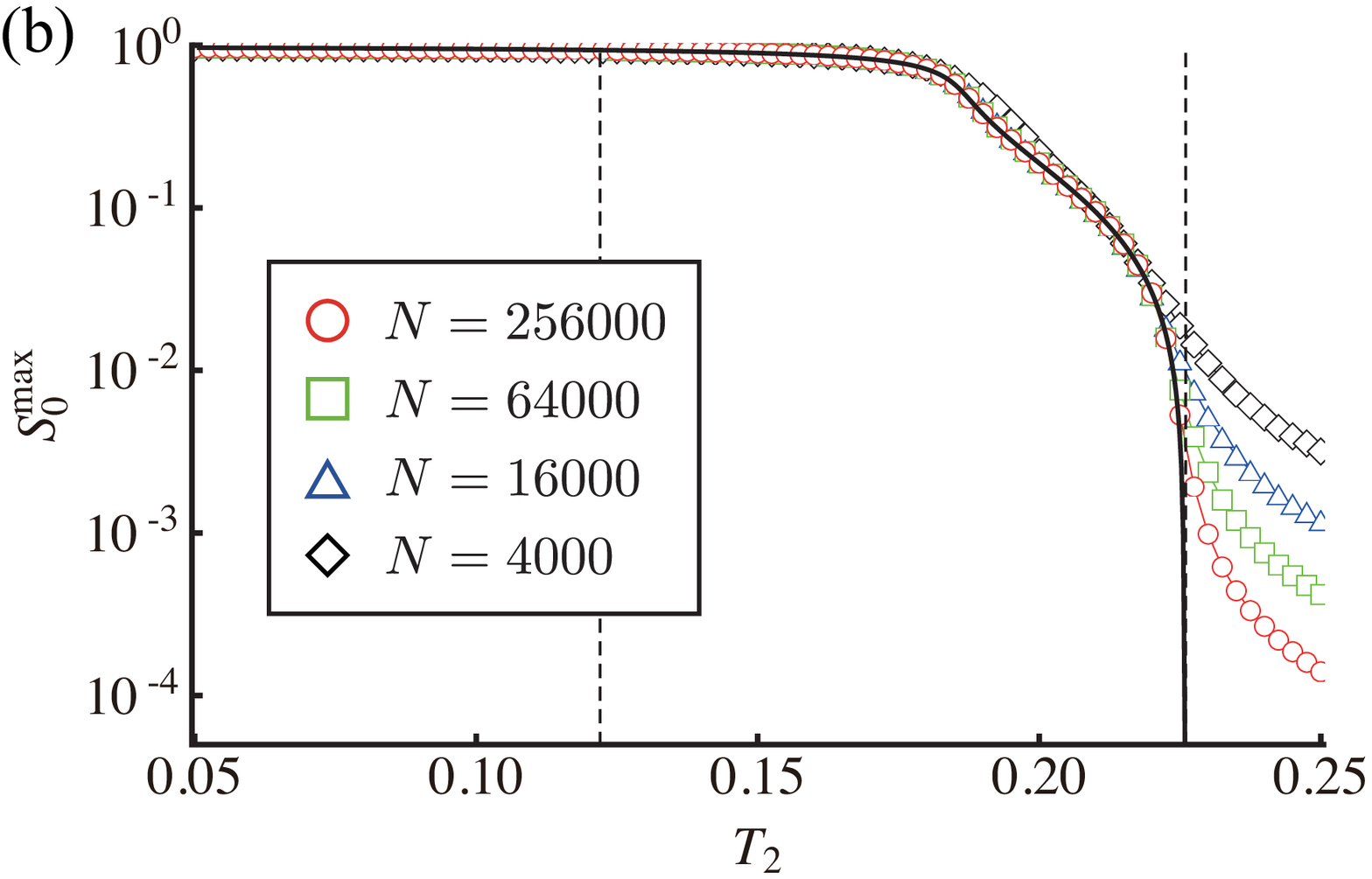}
 \end{center}
 \caption{
Monte--Carlo results for (a) $S_3^{\rm max}$ and (b) $S_0^{\rm max}$ as functions of $T_2$, with $T_1=0.25$ and $\rho=0.01$.
The two vertical lines represents $\tc \simeq 0.122$ and $\tcp \simeq 0.226$.
Here, $\tc$ is given by the cross point of the fractal exponents of differently sized systems. 
The solid line in (b) is drawn by equation (\ref{s0max}), which gives $\tcp$.
Note the lack of any discontinuous jump of $S_3$ (figure~\ref{fig-phasediagramhroneq0}).
}
\label{fig-SmallT1}
\end{figure}

\subsection{Results for random regular network}

We now discuss the sensitivity of the phase diagram to the fraction of initial adopters on a degree-6 random regular network.
Here, we verified our analytical results by Monte--Carlo simulations of the present model.
The number of nodes used is $N=$4000, 16000, 64000, and 256000. 
The number of graph realizations, and also the number of trials on each graph are 100.

Figure~\ref{fig-phasediagramhroneq0} plots the phase diagram when $\rho=0.01$.
The lower and upper bounds ($T_d^L(T_1)$ and $T_d^U(T_1)$) of the discontinuous jump of the $S_3$-network 
and the percolation threshold ($\tcp(T_1)$) of the $S_0$-network are drawn by the tree approximation. 
Moreover, plotted (discrete dots) are the numerically estimated percolation transition points $\tc(T_1)$ of the $S_3$-network, 
where $S_3^{\rm max}$ changes from $S_3^{\rm max}=0$ to $S_3^{\rm max}>0$. 
Each transition point is estimated from the cross point of the effective fractal exponent $\psi(N)$ \cite{hasegawa2013profile}, 
defined as $\psi(N) = {\rm d} \ln N S_3^{\rm max}(N)/{\rm d}\ln N$.
Note that the existence of the crossing point of the effective fractal exponents for for various $N$ indicates a continuous transition \cite{hasegawa2013profile,nogawa2014transition}. 

This phase diagram is crucially different from that of $\rho=0+$.
For $\rho>0$, the upper bound $T_d^U(T_1)$ deviates from the vertical line $T_2=0.2$ (i.e., the discontinuous transition line of the case $\rho = 0+$). 
The continuous transition points $\tc(T_1)$ of the $S_3$-network at small $T_1$ move toward the lower side and never intercept the upper bound $T_d^U(T_1)$ at large $T_1$. 
Then the percolation of the $S_3$-network and the discontinuous jump of $S_3$ occur at different points: 
as $T_2$ increases with a fixed value of $T_1$ ($>$0.294 for $\rho=0.01$), 
a continuous transition of the $S_3$-network occurs, 
followed by a discontinuous change in $S_3$ as $T_2$ approaches the upper bound $T_d^U(T_1)$.

Figure~\ref{fig-comparisonR} compares the analytical and numerical results for $S_3$ with $T_1=0.4$. 
Both results nearly coincide except near the discontinuous transition point $T_2=T_d^U \simeq 0.159$.
As $T_2$ increases, $S_3$ rapidly changes from the lower to the upper branch, 
the cross-over becomes sharpens and steepens, and the location becomes closer to $T_d^U$ as $N$ increases, 
indicating that the rapid change occurs as a sudden jump at $T_d^U$ in the large size limit. 
The size dependence of apparent crossing points also shows that the crossing point approaches $T_d^U$ as $N$ increases (not shown). 
The fraction of the largest component $S_3^{\rm max}$ also changes discontinuously at $T_d^U$ (figure~\ref{fig-Rmax}). 

Discontinuous transitions never appear at small $T_1$, in either the analytical or numerical solutions
(the case of $T_1 = 0.25$ is plotted in figure~\ref{fig-SmallT1}). 
In this case, the $S_3$-network first percolates ($\tc \simeq 0.122$ for $T_1=0.25$), then continuously increases to $S_3=1$ (figure~\ref{fig-SmallT1}(a)).
Figure~\ref{fig-SmallT1}(b) plots the Monte--Carlo result of $S_0^{\rm max}$ with $T_1=0.25$ and $\rho=0.01$.
The numerically evaluated $S_0^{\rm max}$ are well fitted by equation~(\ref{s0max}), implying that 
the $S_0$-network disintegrates at $\tcp(T_1)(>T_d^U(T_1))$ when $T_1<0.344$, as analytically determined. 
When $T_1>0.344$, the percolation transition of the $S_0$-network occurs simultaneously with a discontinuous jump of $S_3$.

\begin{figure}
 \begin{center}
   \includegraphics[width=75mm]{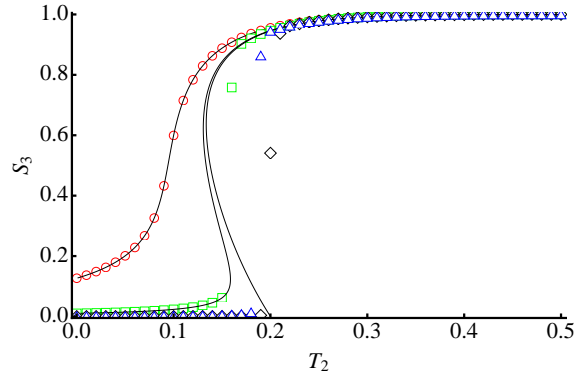}
   \end{center}
 \caption{
Analytically determined $S_3$ as a function of $T_2$ with $T_1=0.4$ and $\rho=0$, 0.01, and 0.1 (right to left).
Symbols are Monte-Carlo results of $N=256000$ and $\rho=$0.1 (red-circles), 0.01 (green-squares), 0.001 (blue-triangles), and 0.0001 (black-diamonds).
}
 \label{fig-analyticR}
\end{figure}

\begin{figure}
 \begin{center}
  \includegraphics[width=75mm]{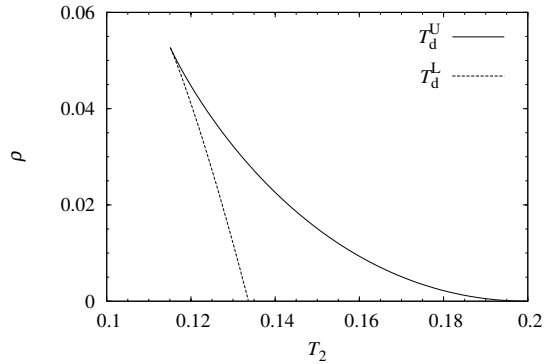}
   \end{center}
 \caption{
Analytically determined lower and upper bounds of discontinuous jumps in the $T_2$--$\rho$ plane with $T_1=0.4$.
}
 \label{fig-criticalrho}
\end{figure}

Finally, analytical curves of $S_3$ versus $T_2$, for fixed $T_1=0.4$ and various values of $\rho$, are plotted in figure~\ref{fig-analyticR}. 
When $\rho$ is small, a precursor of the discontinuous $S_3$ jump is seen, similar to in figure~\ref{fig-analyticalS3rho0}. 
This behavior disappears as $\rho$ increases. 
Figure~\ref{fig-criticalrho} plots the analytically determined lower and upper bounds over a range of $\rho$. 
As noted in the previous subsection, 
a critical fraction $\rho_c \simeq 0.053$ exists at $T_1=0.4$, above which both bounds of discontinuous jump disappear 
and the system exhibits only continuous transitions of percolations of the $S_3$-and $S_0$-networks.

\section{Summary}

We have investigated an MIC model on networks.
Using the tree approximation, we derived formulas for evaluating $S_0$, $S_1$, and $S_3$ and thus constructed the phase diagram. 
As an application, we studied the phase diagram of the present model on several networks, 
and confirmed agreement between the results and our analytical predictions.
In particular, the abandoners discontinuously percolated at large $T_1$. 
Moreover, when the dynamics began from a finite fraction of initial adopters, 
the discontinuous jump in the number of abandoners occurred 
after the percolation of the abandoner. 
It is not until the present MIC model is placed {\it on a network} that we can observe the last behavior. 
As to the original fad model proposed by Krapivsky et al., we find a similar behavior, 
``{\it a discontinuous jump after the percolation}'' when the fad model is placed on a network \cite{hasegawa2014fad}.

Our tree approximation for the present MIC model is valid only on uncorrelated networks. 
Investigating the relationship between the cooperative behavior of the model and the properties of the network 
on which the dynamics is performed (e.g., clustering, degree correlation, and community structure) is beyond the scope of this study. 
For example, clustering decreases the threshold of the SIR model \cite{newman2003properties}, 
but an innovation in the linear threshold model is more easily propagated in clustered networks than in random networks  \cite{ikeda2010cascade,centola2007cascade}.
How the MIC model behaves in clustered networks requires further research.

\ack
This work was partially supported by the Grant-in-Aid for Young Scientists (B) of Japan Society for the Promotion of Science 
(Grant No.~24740054 to T.H.) and JST, ERATO, Kawarabayashi Large Graph Project.

\end{document}